\documentclass[12pt,english,floatfix,superscriptaddress,aps,prd,preprint,showkeys,nofootinbib]{revtex4}
\usepackage{amsmath}
\usepackage{amssymb}
\usepackage{amsbsy}
\usepackage{amsfonts}
\usepackage{amsopn}
\usepackage{amstext}
\usepackage{graphicx}
\usepackage{amssymb}
\usepackage{amsfonts}
\usepackage{amsmath}
\usepackage{graphicx}
\usepackage[english]{babel}
\usepackage{color}
\usepackage{slashed}
\usepackage{esint}
\usepackage[dvips]{epsfig}
\usepackage[dvips]{graphicx}
\usepackage{float}
\usepackage{units}
\usepackage{textcomp}
\newcommand{\beq}{\begin{equation}}
\newcommand{\eeq}{\end{equation}}
\newcommand{\bea}{\begin{eqnarray}}
\newcommand{\eea}{\end{eqnarray}}

\begin{document}

\title{Scalar particles around a Rindler-Schwarzschild wormhole}

\author{C. R. Muniz\footnote{E-mail:celio.muniz@uece.br}}\affiliation{Universidade Estadual do Cear\'{a}, FECLI, Av. D\'{a}rio Rabelo, s/n, 63.500,Iguatu-CE,
Brazil.}
\author{H. R. Christianssen\footnote{E-mail:hugo.christianssen@ifce.ce.br}}\affiliation{Instituto Federal de Ci\^{e}ncias, Educa\c{c}\~{a}o e Tecnologia,
62042-030, IFCE, Brazil.}
\author{M. S. Cunha\footnote{E-mail:marcony.cunha@uece.br}}\affiliation{Universidade Estadual do Cear\'a, CCT, 60714-903, Fortaleza, CE, Brazil.}
\author{J. Furtado\footnote{E-mail:job.furtado@ufca.edu.br}}\affiliation{Universidade Federal do Cariri, Centro de Ci\^encias e Tecnologia, 63048-080,
Juazeiro do Norte, CE, Brasil.}
\author{V. B. Bezerra\footnote{E-mail:valdir@fisica.ufpb.br}}\affiliation{Departamento de F\'{i}sica, CCEN, Universidade Federal da Para\'{i}ba, Caixa Postal
5008, Jo\~{a}o Pessoa, PB, Brazil}
\begin{abstract}
In this paper, we study quantum relativistic features of a scalar field around the Rindler-Schwarzschild wormhole.
First, we introduce this new class of spacetime, investigating some energy conditions and
verifying their violation in a region nearby the wormhole throat,
which means that the object has to have an exotic energy in order to prevent its collapse.
Then, we study the behavior of the massless scalar field in this spacetime
and compute the effective potential by means of tortoise coordinates.
We show that such a potential is attractive nearby the wormhole throat and that is traversable via quantum tunneling by massive
particles with sufficiently low energies.
The solution of the Klein-Gordon equation is obtained subsequently, showing that the energy spectrum of the field is subject to a constraint
which induces a decreasing oscillatory behavior.
On imposing Dirichlet boundary conditions on a spherical shell nearby the throat we then determine the particle energy levels, and we use this spectrum
to calculate the quantum revival of the eigenstates.
Finally, we compute the Casimir energy associated with the massless scalar field at zero temperature.
We perform this calculation by means of the sum of modes method. The zero-point energy is regularized using the Epstein-Hurwitz zeta-function.
We also obtain an analytical expression for the Casimir force acting on the shell.
\end{abstract}


\renewcommand{\thesection}      {\Roman{section}}

\maketitle
\newpage
\section{Introduction}
Wormholes are hypothetical topologically non-trivial spacetime objects connecting arbitrarily distant regions of the universe, or even different universes. These structures, although not predicted by general relativity (GR), are viable solutions of Einstein equations in GR and in many extended theories of gravity.
The term wormhole, in fact, was coined \cite{MW1957} more than two decades after the first such vacuum solution, the so-called Einstein-Rosen bridge \cite{ER1935}, was found.
Later on, it was shown that this type of wormhole is unstable if it connects two parts of the same universe, collapsing off too quickly for any particle (even light) to pass from one side to the other \cite{HW1962}.
Solutions corresponding to traversable wormholes were found in 1973 by Bronnikov and
Ellis \cite{B1973,E1973} and became well-known after the papers of Morris, Thorne and Visser \cite{MT1988,V1989}.

The throat of a wormhole is in principle unstable to small perturbations and collapses under its own mass gravity.
%
%
In brane models, negative pressure can be obtained from the brane tension. Such models assume that our universe is a (3 + 1)-dimensional subspace embedded in a higher-dimensional bulk space. The brane can bend hard enough to make space points far apart on the brane be close in the bulk and thereafter be potentially connected through a wormhole.
If we assume the presence of a BH on each side, against their gravitational attraction there will be the repulsion between folds due to the brane tension. Brane tension thus compensates the demand for exotic matter \cite{DMS2020}.
Another way to balance gravity is to place the structure in a de Sitter background \cite{DMS2017}.

Traversable wormholes generally require the input of exotic sources of negative energy in order to avoid the collapse \cite{SH2002}. In extended theories of gravity wormholes may exist even without exotic matter and even without matter, see for instance \cite{BK2019, GW2007, RS2007, ERS2007}.  Fortunately, an arbitrarily thin layer of negative energy density is sufficient to keep a macroscopic wormhole stable \cite{VKD2003}. Negative energy density can be for instance  produced by Casimir forces \cite{casimir}. Indeed, the Casimir effect has been proved to occur among ordinary matter pieces of diverse shapes and materials \cite{casimir2}. The nature of this effect, as originally pointed out, is connected with the zero-point energy  of the quantum (electrodynamical) vacuum distorted by parallel plates and depends strongly on the geometry of the boundaries. In \cite{MTY1988} it was shown that the plates separation should be smaller than the electron Compton wavelength so that the physical device cannot be realized. Later,  it was concluded that the mass of the plates compensate the negative energy density and the traversable wormhole is not completed \cite{Visser1995, FR1996}. Recently, Garattini \cite{Garattini2019}  called attention to the fact that nothing was said about the possible forms of the shape function and the redshift function in the literature. These functions are related when imposing an equation of state among radial pressure and energy density $p_r= \varpi \rho$. For $ \varpi=3$ one meets the conserved Casimir stress energy tensor (contributing only at the Planck scale); for $\varpi=1$ the Ellis-Bronnikov wormhole (of sub-planckian size) and
$ \varpi=-1$ is for phantom energy.

The actual existence of traversable wormholes is allowed in GR if quantum effects are included.
Indeed, during the quantum gravity phase of the primordial universe, large quantum fluctuations in the geometry and topology of spacetime could likely give rise to such structures and provide the negative energy density necessary to stabilize them. Although microscopic, such primordial quantum configurations could be made macroscopic during expansion and thus potentially observable \cite{DGV2017}. The potential existence of astrophysical wormholes is discussed in \cite{BS2021}.

In the next section, we introduce the the Rindler-Schwarzschild wormhole metric that we will consider throughout this paper and we discuss the violation of Null and Average Null Energy Conditions around it. In Sect.\ref{sect:scalars} we analyze the equation of motion of a scalar field surrounding the wormhole spacetime and obtain its spectrum and solutions considering appropriated boundary conditions. In Sect.\ref{sect:revival} we compute the quantum revival of the wave function. Section \ref{sect:casimir} is dedicated to the calculation of the Casimir energy of the scalar field in the wormhole for a thin spherical shell close the throat. In the final Section we draw our conclusions.

\section{Metric and Energy Conditions}\label{sect:metric}
The first wormhole solution was originally constructed after the static black hole spherically symmetric Schwarzschild  solution \cite{ER1935}.
This metric contains two asymptotically flat spacetimes, $r > 2GM$  and $r < 2GM$. These are two disconnected spacetimes sharing the same horizon at the $r = 2GM$ hypersurface. The existence of an event horizon
prevents such a wormhole from being traversable.  One makes it traversable by copying the space given by $r > 2GM$ and then pasting the two copies together. In this representation, the wormhole does not connect separate points of the same universe but two different universes each one with its own black hole, infinitely distant from the other.

We introduce a novel wormhole spacetime which asymptotically behaves as a non-flat Rindler-like spacetime given by
\begin{equation}\label{MRLS}
ds^2=\alpha^q r^q dt^2-\frac{dr^2}{1-b(r)/r}-r^2d\Omega^2,
\end{equation}
where $b(r)$ is the shape function. The usual Rindler spacetime, recovered for $q=2$ and large $r$, is the coordinate frame of an observer undergoing constant proper acceleration in an otherwise flat Minkowski spacetime \cite{R1966}. Note that while we assume a Schwarzschild-like $g_{rr}$ we replace $g_{tt}$ by a Rindler type function.
For $b(r)=2GM\equiv b_0$ this spacetime has a Ricci scalar curvature given by
\begin{equation}
R_S=-\frac{q [q (r-b_0)+b_0-2 r]}{2 r^3}.
\end{equation}
Differently from the Rindler-Ellis-Bronnikov wormhole \cite{Kar}, this curvature depends manifestly on the throat radius, except for $q=1$.

The Null Energy Conditions (NEC) for the perfect fluid that generate
the Rindler-Schwarzschild wormhole are violated in a finite region nearby the throat. Indeed,
\begin{eqnarray}
\rho &+&p_r =\frac{2r-3b_0}{r^3}< 0\ \text{in the interval}\ b_0 \leq r<\frac{3}{2}b_0, \nonumber\\
\rho &+&p_{\phi}(=p_{\theta})=\frac{1}{r^2}\geq 0\ \text{for all}\ r,
\end{eqnarray}
with $\rho$ and $p_r$ being the energy density and radial pressure, respectively, and $p_{\phi}$, $p_{\theta}$ the lateral pressures.
Such a violation in NEC occurs in a region slightly larger than the one corresponding to the deformed Ellis-Bronnikov wormhole \cite{Kar}. Taking into account the generalized Rindler-Ellis-Bronnikov wormhole, with shape function given by $b(r)=r-r^{3-2m}(r^m-b_0^m)^{2-2/m}$  ($m\geq2$ is an even integer \cite{KarII}, noting that $m=2$ yields the usual Ellis-Bronnikov wormhole), we have that for $m\geq4$ the referred violation occurs in a region even smaller since this region now corresponds to the interval $b_0\leq r< m^{1/m}b_0$.

Regarding the Averaged Null Energy Conditions (ANEC) $\int T_{\mu\nu}k^{\mu}k^{\nu}d\lambda\geq0$, there is again a violation.
ANEC must be calculated along radial null geodesics and result in
\begin{equation}
\frac{1}{\kappa\alpha}\int_{b_0}^{\infty}\frac{2r-3b_0}{r^3\sqrt{r(r-b_0)}}dr=-\frac{8}{15\kappa\alpha b_0^2}.
\end{equation}
Note however that, fixing $\alpha$, this violation can be neglected for large wormhole throats ({\it i.e.}, for macroscopic deformed Schwarzschild-like wormholes). Actually, such a violation is smaller than the one obtained for the wormhole discussed in \cite{Kar}.

\section{Massless scalar field and energy levels}\label{sect:scalars}
We will consider the radial part of the Klein-Gordon equation, $\frac{1}{\sqrt{-g}}\partial_{\mu}[\sqrt{-g}g^{\mu\nu}\partial_{\nu}\phi]+m^2\phi=0$, for a
scalar field with profile $\phi(r,\theta,\varphi,t)=\exp{(-i\omega t)}Y_{\ell}^{k}(\theta,\varphi)R(r)$ ($k$ is the azimuthal quantum number) placed in the
spacetime given by Eq. \ref{MRLS}. Our main focus here will be on the massless field but for the sake of completeness we will also consider massive particles in some calculations. Performing the transformation $R(r)=\xi(r)\sqrt{g_{rr}}/r$, we obtain an equation for
$\xi(r)$ which turns to be a Schr\"{o}dinger-type equation on introducing the tortoise radial coordinate, $r_*$, as follows
\begin{equation}
\frac{dr_*}{dr}=\sqrt{-\frac{g_{rr}}{g_{00}}}=\alpha^{-q/2}r^{-q/2}\left(1-\frac{b_0}{r}\right)^{-1/2}.
\end{equation}
For $q=2$ we have
\begin{equation}
r=\frac{1}{4} b_0e^{-\alpha r_*} \left(1+e^{\alpha r_*}\right)^2,
\end{equation}
so that at the wormhole throat, $r=b_0$, $r_*=0$.
The Schr\"{o}dinger-type equation is given by,
\begin{equation}\label{Schr-likeEq}
-\frac{d^2\xi(r)}{dr_*^2}+\left[V_{eff}(r) -\omega^2\right]\xi(r) =0,
\end{equation}
where the effective potential is
\begin{equation}
V_{eff}=\frac{\alpha^2 \left\{b_0^2+4 b_0 \left[(2 \ell+1)^2 r-4 m^2 r^3\right]-4 r^2 \left[ 4\ell(\ell+1)-4 m^2 r^2+1\right]\right\}}{16 r (b_0-r)},
\end{equation}

\begin{figure}[h!]
    \centering
            \includegraphics[width=0.6\textwidth]{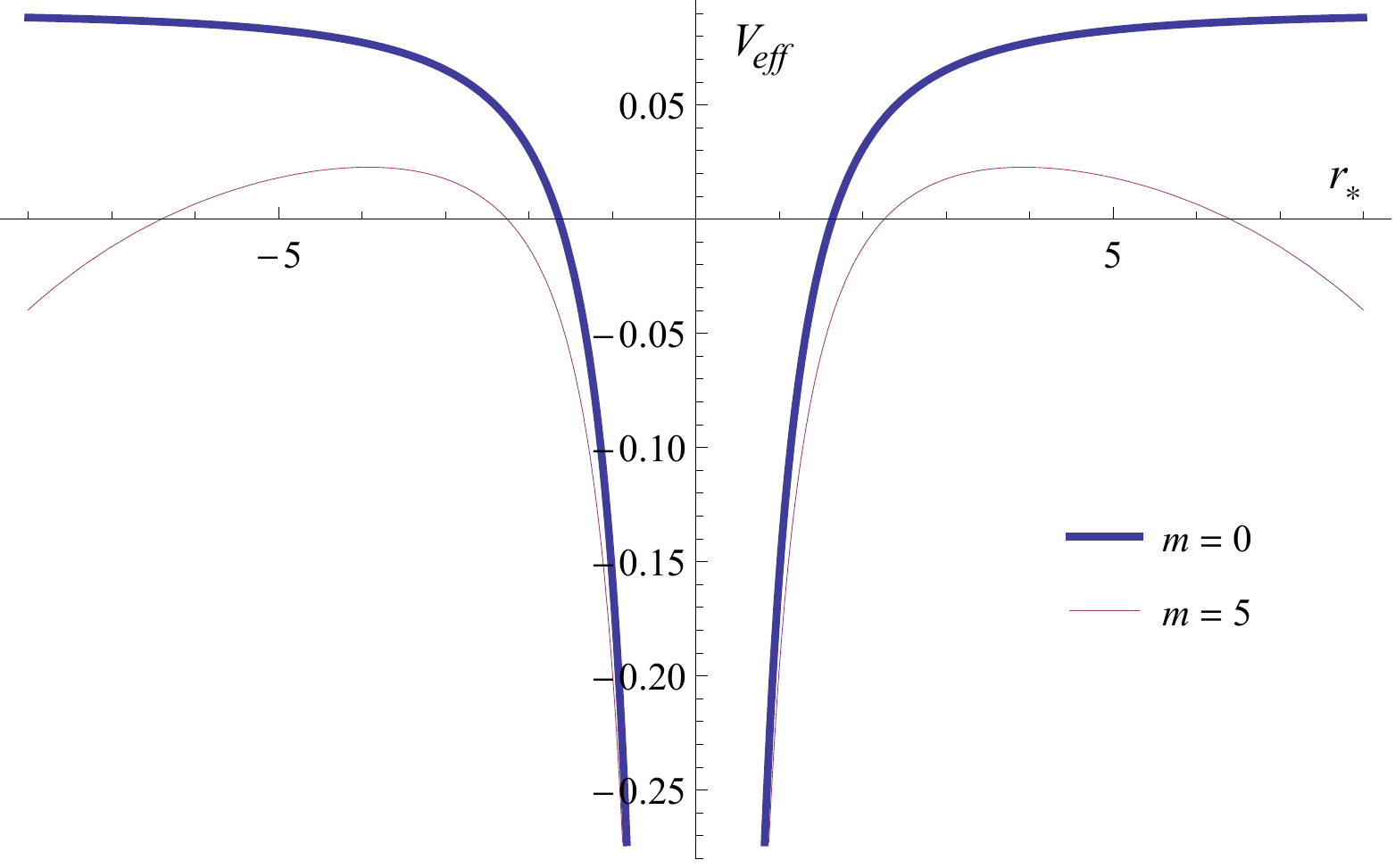}
   \caption{Effective potential for both massive and massless scalar fields around the Rindler-Schwarzschild wormhole ($q=2$), with $\alpha=0.2$, $\ell=1$ and $b_0=0.2$ in Planck units.}
        \label{BSDMFig0}
\end{figure}

In Fig. \ref{BSDMFig0}, we show the potential resulting from the Rindler-Schwarzschild wormhole in terms of the tortoise
coordinate for both massive and massless scalar particles. Nearby the wormhole throat, the effective potential is attractive and falls as $-1/r_*^2$. Notice that for massive particles with low energies, at sufficient distances from the throat there is the possibility of the particle to traverse the wormhole via quantum tunnelling.

\begin{figure}[!ht]
    \centering
    \begin{minipage}{0.5\linewidth}
        \centering
        \includegraphics[width=0.95\textwidth]{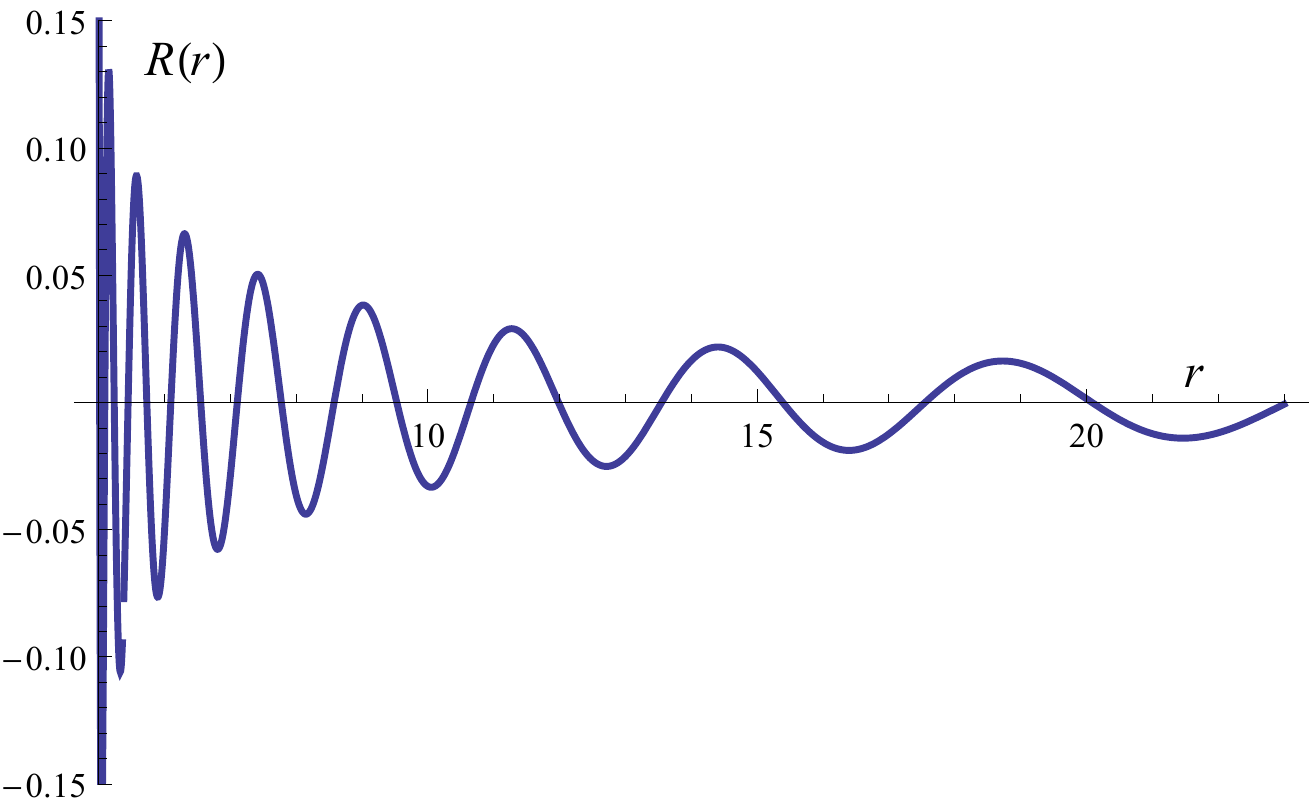}
        \label{fig:figura1minipg}
    \end{minipage}\hfill
    \begin{minipage}{0.5\linewidth}
        \centering
        \includegraphics[width=0.95\textwidth]{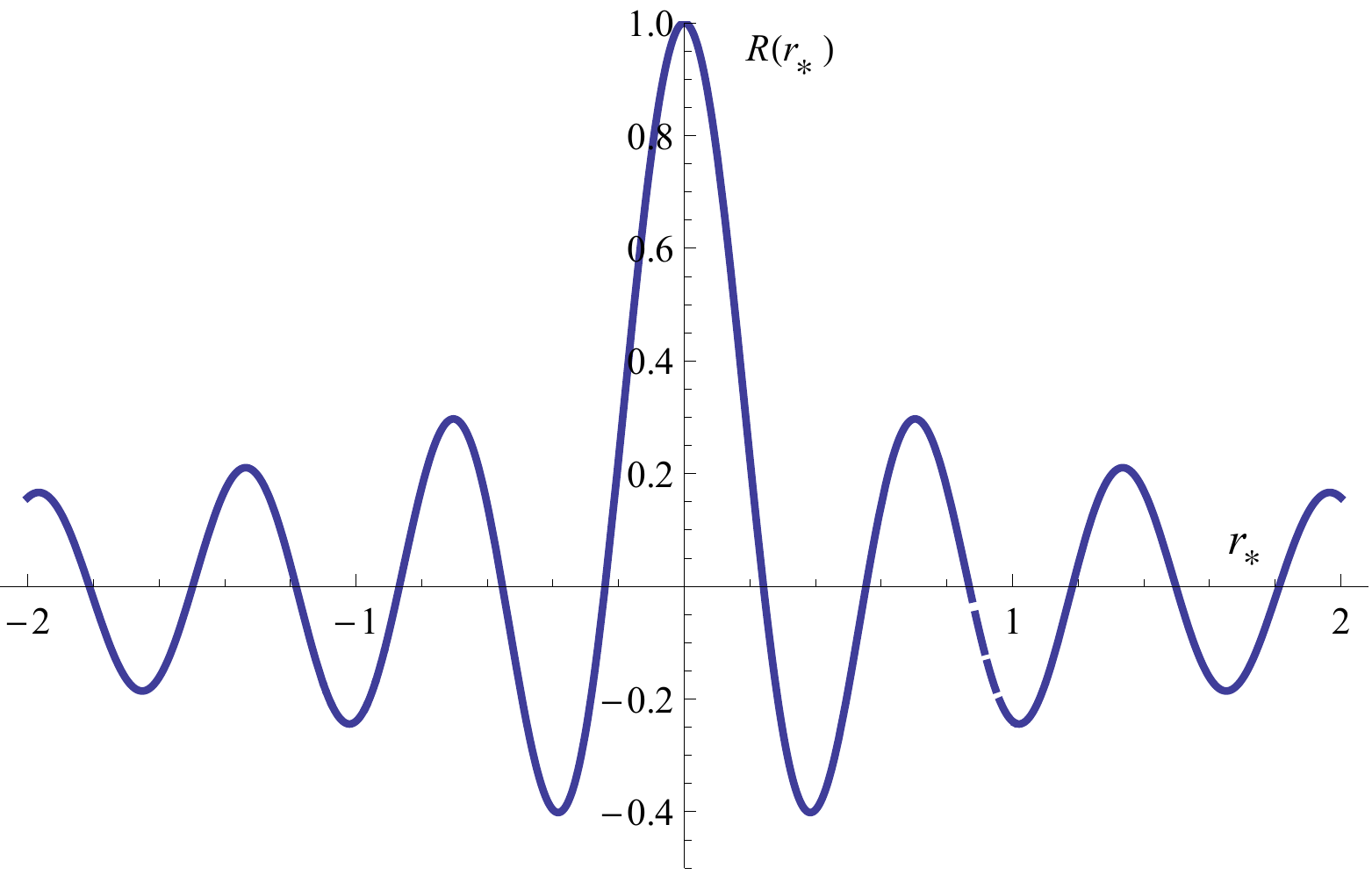}
              \label{fig:figura2minipg}
    \end{minipage}
   \caption{Radial part of the massless scalar field around the deformed Rindler-Schwarzschild wormhole, in terms of the radial coordinate (left), and tortoise coordinate (right). The parameter settings are $\ell=1$, $b_0=1$, $\alpha=0.5$, and $\omega=10$ in Planck units.}
    \label{BSDMFig1}
\end{figure}

The solutions of Eq. (\ref{Schr-likeEq}) in the Rindler-Schwarzschild wormhole are given in terms of the Legendre functions of first ($P_s(z)$) and second
($Q_s(z)$) types, as follows
\begin{equation}
R(r)=C_1 P_s\left(\frac{2 r}{b_0}-1\right)+C_2 Q_s\left(\frac{2 r}{b_0}-1\right),
\end{equation}
where $s={\frac{\sqrt{\alpha^2(2 \ell +1)^2-4 \omega^2}-\alpha}{2 \alpha}}$. Note that $Q_s\left(\frac{2 r}{b_0}-1\right)$ diverges at the throat, $r=b_0$, and then we will take $C_2=0$. In Fig. \ref{BSDMFig1} we depict the behavior of the $R(r)$ as a function of the radial (left panel) and tortoise (right panel) coordinates, for $C_1=1$.

The decreasing oscillatory behavior of the radial function corresponds to a complex $s$ which is given by $s=-\frac{1}{2}+i\frac{\sqrt{4
\omega^2-\alpha^2(2 \ell +1)^2}}{2 \alpha}$. This implies the following constraint on the particle energy
\begin{equation}
\omega>\left| \frac{\alpha(2\ell+1)}{2}\right|.
\end{equation}
We draw attention to the fact that such a behavior of the massless scalar field is the same discussed in \cite{Kar} for the deformed Ellis-Bronnikov wormhole.

 We consider now the hypergeometric representation of $P_s(z)$, given by
\begin{equation}
P_s(z)=\, _2F_1(-s;s+1;1;\frac{1-z}{2}),
\end{equation}
valid in the interval $1<z<\infty$. Thus, our radial solution becomes
\begin{equation}\label{FirstHyper}
R(r)=C_1\,_2F_1\left(\frac{1}{2}-i\lambda;-\frac{1}{2}+i\lambda;1;1-\frac{r}{b_0}\right),
\end{equation}
where we have defined the dimensionless quantity $\lambda=\sqrt{4\omega^2-\alpha^2(2\ell+1)^2}/2\alpha$. In order to compute
some possible energy eigenvalues for a particle in orbit around the wormhole, let us expand Eq. (\ref{FirstHyper})
around $r=b_0$, up to first order in $(1-r/b_0)$. In this region, we find
\begin{equation}
R(r)\propto1+\left(\lambda^2+\frac{1}{4}\right)\left(1-\frac{r}{b_0}\right)+\mathcal{O}\left[\left(1-\frac{r}{b_0}\right)^2\right].
\end{equation}
From now on, we will use a ``hard-shell condition'', so that nearby the wormhole throat, $r=r_s\approx b_0$, we get $R(r_s)\approx 0$.
In fact, for arbitrary high energies, the potential barrier profile approximates the wormhole throat and the energy levels are given by
\begin{equation}\label{EnergyWallRadius}
\omega_{\ell}\approx \alpha\sqrt{\ell(\ell+1)+\frac{b_0}{r_s-b_0}},
\end{equation}
which are higher the closer is the shell to the wormhole throat.

\section{Quantum revival characteristic time}\label{sect:revival}

Quantum revival is said to occur when the wave function recovers its initial state at an instant dubbed revival time.
For a particular quantum number $\nu_i$ it is defined by
\begin{equation}
\tau=\frac{4\pi}{|\left(\frac{\partial^2 \omega_{\nu}}{\partial \nu^2}\right)_{\nu=\nu_i}|}.
\end{equation}
\begin{figure}[h!]
    \centering
            \includegraphics[width=0.50\textwidth]{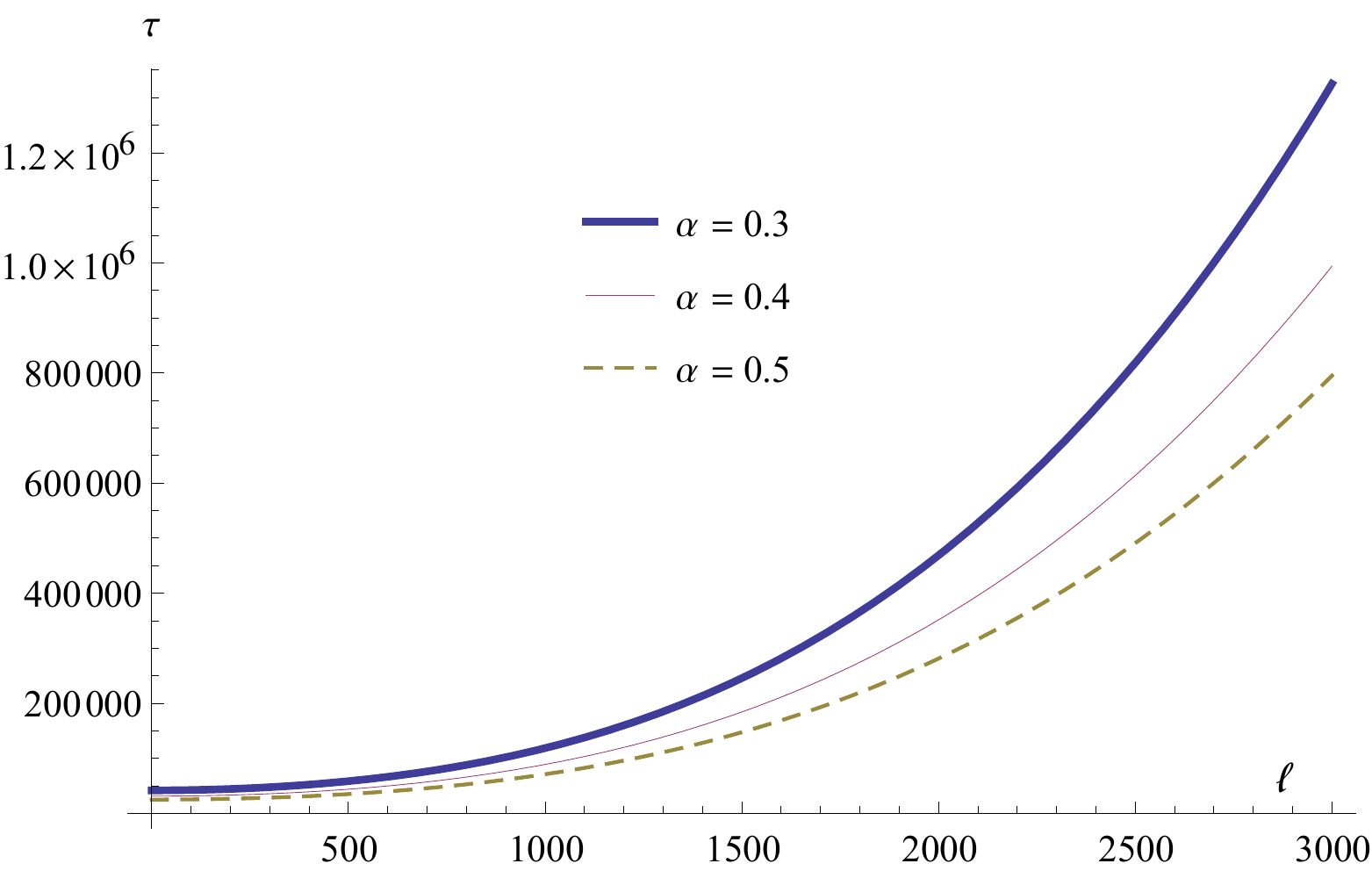}
    \caption{Quantum revival time of the eigenstates of the massless scalar field around the deformed Rindler-Schwarzschild wormhole given in terms of the polar quantum number $\ell$, for some values of $\alpha$, with $b_0=1.0$ and $r_s=1.000001 b_0$, in Planck units.}
    \label{Tunnel-2-1}
\end{figure}
The energy levels obtained in Eq.(\ref{EnergyWallRadius}) depend on just one quantum number, the polar quantum number $\ell$. Thus, for the $\ell$ eigenstate the quantum revival is given by
\begin{equation}
\tau\approx \left|\frac{16 \pi \left[(b_0-r_s)\ell (\ell+1) -b_0\right] \sqrt{\frac{b_0}{r_s-b_0}+\ell(\ell+1)}}{ \alpha  (r_s-5 b_0)}\right|.
\end{equation}
In Fig. \ref{Tunnel-2-1}, we depict these characteristic revival times as a function of $\ell$ for some values of the deformation parameter, $\alpha$. Notice that the minimum occurs for the null angular momentum, namely, $\ell=0$, yielding
\begin{equation}
\tau_{min}\approx\left|\frac{16\pi b_0}{\alpha(r_s-5b_0)}\sqrt{\frac{b_0}{r_s-b_0}}\right|.
\end{equation}

\section{Casimir energy and force}\label{sect:casimir}

In this section, let us calculate the Casimir energy $\mathcal{E}_C$ of the massless scalar field around the deformed Rindler-Schwarzschild wormhole in the presence of a thin spherical shell nearby the throat. We will do it through a direct summation of modes. As we have seen previously, the field satisfies the Dirichlet boundary conditions on the shell.
We shall adopt a suitable procedure of regularization in order to obtain a finite result.
Considering Eq. (\ref{EnergyWallRadius}) and using the fact that $r_s\approx b_0$,
the initial expression for the zero-point energy for the scalar field is given by
\begin{equation}\label{ZeroPoint}
E_0=\frac{1}{2}\sum_{\ell=0}^{\infty}g(\ell)\omega_{\ell}=\alpha\sum_{\ell=0}^{\infty}\left(\ell+\frac{1}{2}\right)\sqrt{\left(\ell+\frac{1}{2}\right)^2+\frac{b_0}{r_s-b_0}-\frac{1}{4}},
\end{equation}
\begin{figure}[b!]
    \centering
            \includegraphics[width=0.65\textwidth]{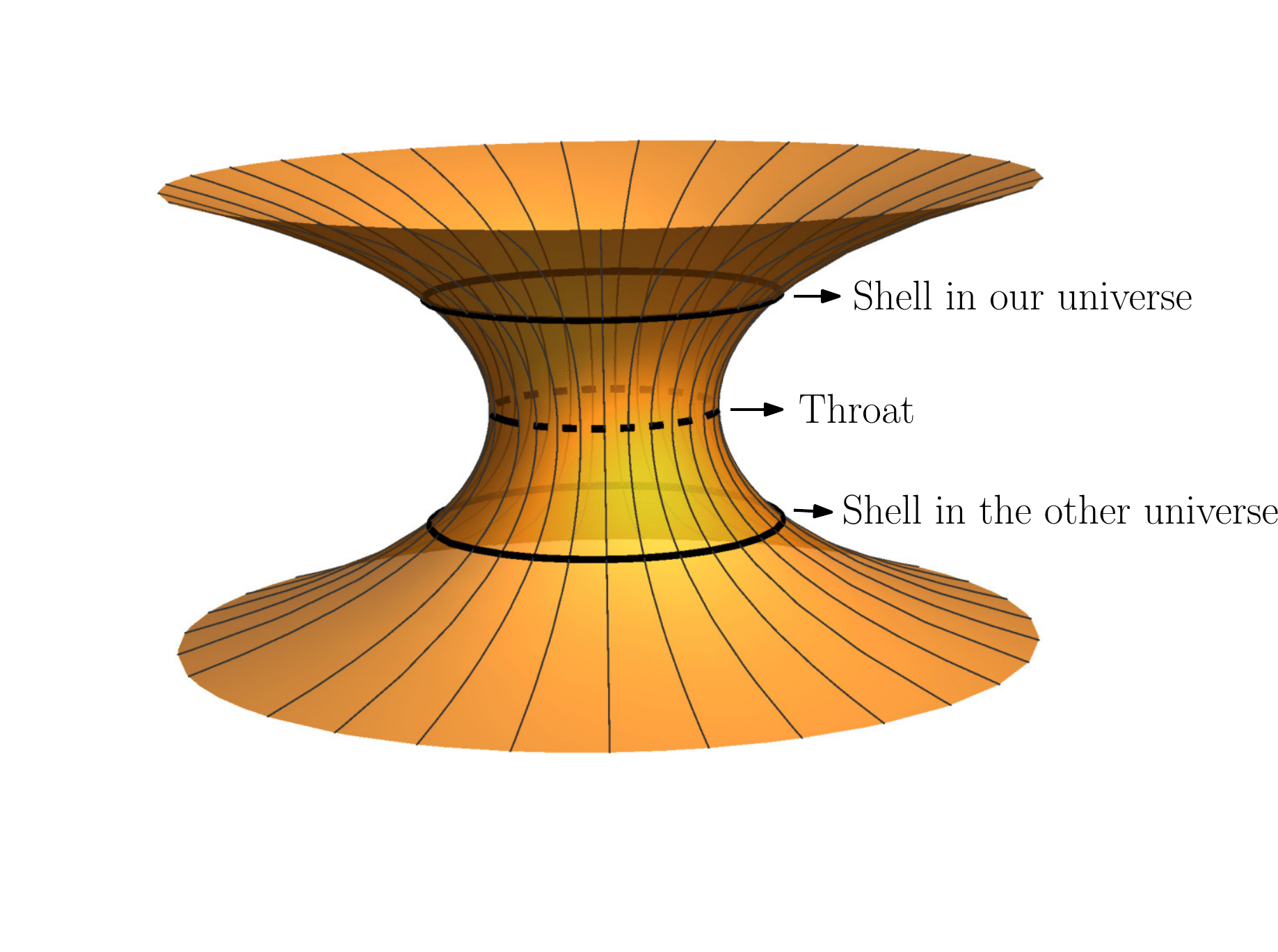}
        \caption{Immersion diagram of the deformed Schwarzschild-like wormhole, with the thin material shells nearby the throat at its two opposite sides.}
    \label{Tunnel-2-3}
\end{figure}
where $g(\ell)=2\ell+1$ is the system degeneracy.

In order to regularize the (divergent) zero-point energy we will employ the zeta-function procedure.
Before doing this, let us express $E_0$ in the form \cite{Elizalde0}
\begin{equation}\label{E-HDerivative}
E_0=\frac{\omega_0}{2}+\frac{\alpha}{1-s}\frac{\partial \zeta_{EH}(s-1;a,b)}{\partial a},
\end{equation}
where $\zeta_{EH}(s;a,b)$ is the Epstein-Hurwitz zeta-function given by $\zeta_{EH}(s;a,b)=\sum_{n=1}^{\infty}[(n+a)^2+b]^{-s}$, with $b>0$.
In the present case we have $s=-1/2$, $a=1/2$, and $b=b_0/(r_s-b_0)-1/4\approx b_0/(r_s-b_0)$.
The regularization can be seen by using the identity \cite{Elizalde}
\begin{eqnarray}\label{EHFormula}
\zeta_{EH}(s;a,b)&=&\frac{b^{-s}}{\Gamma{(s)}}\sum_{n=0}^{\infty}\frac{(-1)^n\Gamma{(n+s)}}{n!}b^{-n}\zeta{(-2n,a)}+\frac{\sqrt{\pi}\Gamma{(s-1/2)}}{2\Gamma{(s)}}b^{1/2-s}\nonumber\\
&+&\frac{2\pi^s}{\Gamma{(s)}}b^{1/4-s/2}\sum_{n=1}^{\infty}n^{s-1/2}\cos{(2\pi n a)}K_{s-1/2}(2\pi n \sqrt{b}),
\end{eqnarray}
where $\zeta(p,q)$ is the Hurwitz zeta-function and $K_{\mu}(z)$ is the Bessel function of second type (Macdonald's function).
 Notice that the second term of the r.h.s of Eq. (\ref{EHFormula}) contains an explicit divergence in the factor $\Gamma{(s-1/2)}=\Gamma{(-1)}$ \cite{Nester,Herondy1}. However, the derivative in Eq. (\ref{E-HDerivative}) removes this issue. Thus, using Eq. (\ref{EHFormula})
to compute Eq. (\ref{E-HDerivative}), the  Casimir energy results in
\begin{equation}
\mathcal{E}_C\approx \frac{\alpha}{2}\sqrt{\frac{b_0}{r_s-b_0}}\left[1+\frac{3}{4\sqrt{\pi}}\sum_{n=0}^{\infty}\frac{(-1)^n n
\Gamma(n-3/2)}{n!}\left(\frac{b_0}{r_s-b_0}\right)^{1-n}(-1+2^{1-2n})\zeta(1-2n)\right],
\end{equation}
where the factor outside the brackets is related to the zero-mode energy $\omega_0$, and $\zeta[z]$ is the Riemann zeta-function. The last term of Eq. (\ref{EHFormula}) can be neglected since the Macdonald's function behaves asymptotically as a decreasing exponential of the argument
(recall that $b_0/(r_s-b_0)\gg 1$).
\begin{figure}[h!]
    \centering
            \includegraphics[width=0.55\textwidth]{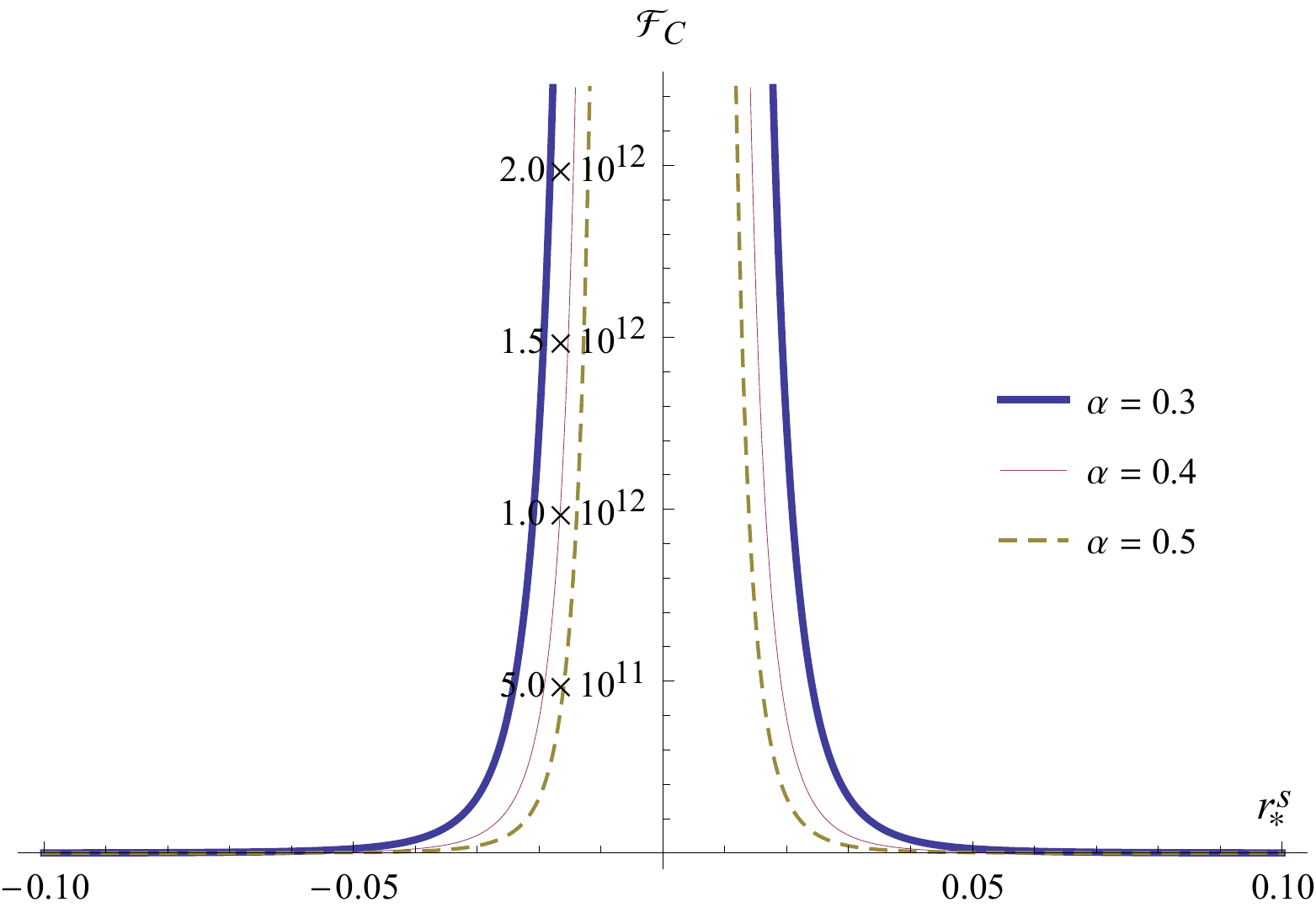}
        \caption{Casimir force on thin material shells nearby the throat of a Rindler-Schwarzschild wormhole at its two opposite sides. The
        force is due to the vacuum of the massless scalar field obeying the Dirichlet boundary conditions, here depicted as a function of the shell
        radius in tortoise coordinates, $r_{*}^s$, for some values of $\alpha$ and for $b_0=1$, in Planck units.}
    \label{Tunnel-2-4}
\end{figure}
Fig. \ref{Tunnel-2-4} shows the behavior of the Casimir force on each shell situated at symmetrical opposite sides nearby the wormhole throat (see the immersion diagram of Fig. \ref{Tunnel-2-3}).
The referred force can be computed as $\mathcal{F}_C=-\partial \mathcal{E}_C/\partial r_s$ and it
is depicted as a function of the shell radius in tortoise coordinates, $r_*^s$. Notice that the force tends to expand the shells outwards the wormhole throat. The Casimir force is bigger the bigger as the parameter $\alpha$. On the other hand, the closer is the shell to the throat the stronger is this force. In fact, when $r_s\to b_0$ we have that $\mathcal{E}_C\approx (\alpha/4)\sqrt{b_0/(r_s-b_0)}$ and the following result arises
\begin{equation}
\mathcal{E}_C\approx \frac{2\alpha}{3}\left(\frac{b_0}{r_s-b_0}\right)^{3/2}+\frac{13\alpha}{24}\left(\frac{b_0}{r_s-b_0}\right)^{1/2},
\end{equation}
and the corresponding Casimir force reads
\begin{equation}
\mathcal{F}_C\approx\frac{\alpha e^{-\alpha r_*^s}}{6 b_0} \left[\frac{e^{\alpha r_*^s}}{\left(e^{\alpha r_*^s}-1\right)^2}\right]^{5/2} \left(166 e^{\alpha r_*^s}+13 e^{2 \alpha r_*^s}+13\right),
\end{equation}
in tortoise coordinates.


\section{Conclusion}

In this paper we have studied quantum relativistic features of a scalar field around the Rindler-Schwarzschild wormhole. Firstly, we presented this novel type of spacetime, which is asymptotically a Rindler spacetime. Then we investigated NEC and ANEC, verifying that these energy conditions are violated in a region nearby the wormhole throat, allowing us comparing these conditions with the ones obtained in \cite{Kar} with respect to the Rindler-Ellis-Bronnikov wormhole. Regarding the NEC, we also have related our findings with those ones associated with the generalized Ellis-Bronnikov solution \cite{KarII}, where the violation region is even smaller. Thus, we conclude that our wormhole solution needs an exotic matter too, in order to prevent it from collapsing.

The study of the behavior of the massless scalar field around this spacetime permitted us finding the effective potential via tortoise coordinates transformation. We have shown that such a potential is attractive nearby the wormhole throat, and massive particles with low energies can quantum tunnel through it. Then we obtained the solution of the Klein-Gordon equation, showing that the energy of the field is subject to a constraint in order to present a decreasing oscillatory behavior. On imposing Dirichlet boundary conditions (b.c.) on a spherical shell nearby the throat, we determine the particle energy levels. Taking these data into account, we have calculated the quantum revival time, in other words, the minimum time necessary to the field return to this initial state. As a conclusion, we can say that this time decreases with the increasing of the deformation parameter, $\alpha$, and it increases with the increasing of the quantum number $\ell$ of the particle.

Finally, we have computed the Casimir energy associated to the massless scalar field at zero temperature, by means of the sum of modes method, taking into account the degeneracy of the system. On regularizing the obtained zero-point energy via Epstein-Hurwitz zeta function, we have arrived to the Casimir energy, and then obtained the related force, depicted in the Fig. \ref{Tunnel-2-4} as a function of the shell positions, in tortoise coordinates. Then, we can see the force on the spherical shells placed on both sides of the throat, which tends to radially expand them, is bigger the smaller as the deformation parameter, $\alpha$. We can also notice that such a force is more intense the closer the shells are to the throat; this feature is consistent with the fact that the presence of an exotic energy, as the Casimir one, prevents the wormhole throat from collapsing.

\section*{Acknowledgements}
The authors thank the Conselho Nacional de Desenvolvimento Cient\'{i}fico e Tecnol\'{o}gico (CNPq), grants n$^o$ 308268/2021-6 (CRM), 307211/2020-7 (VBB), and 315926/2021-0 (MSC) for financial support.



\end{document}